\documentclass[11pt,dvips]{article}

\usepackage{epsfig,times} 
\usepackage{picinpar}
\usepackage{wrapfig}
\usepackage{floatflt}

\makeatletter
\renewcommand{\section}{\@startsection{section}{1}{0in}
	{0.4\baselineskip}{0.1\baselineskip}{\Large\bf}}
\renewcommand{\subsection}{\@startsection{subsection}{2}{0in}
	{0.25\baselineskip}{-\baselineskip}{\large\bf}}
\renewcommand{\subsubsection}{\@startsection{subsubsection}{3}{0in}
	{0.1\baselineskip}{-\baselineskip}{\normalsize\bf}}
\makeatother
\newcommand{\icrc}{$26^{\rm th}$ ICRC\ (Salt Lake City, 1999)}
\def\gray{$\gamma$-ray\ }

\def\BC{B/C}
\def\Berat{$^{10}$Be$/\,^9$Be}
\def\Alrat{$^{26}$Al$/\,^{27}$Al}
\def\Clrat{$^{36}$Cl$/\,$Cl}
\def\Mnrat{$^{54}$Mn$/\,$Mn}
\def\Ferat{sub-Fe/Fe}

\def\Dunits{$\times10^{28}$ cm$^2$ s$^{-1}$}

\def\vunits{km s$^{-1}$}

\newcommand{\aap}{A\&A\ }
\newcommand{\apj}{ApJ\ }

\def\ocircle{~\put(0,2){\tiny $\bigcirc$}\hspace*{1em}}
\def\obox{~\put(0,4){\fbox{}}\hspace*{1em} }
\def\fcircle{{\Large $\bullet$} } 
\def\odiamond{{\Large $\diamond$} }
\def\otriangle{{\footnotesize $\triangle$} }
\def\fha{110mm}
\def\fwa{169mm}
\def\fwb{84mm}

\begin{document}

%
\makeatletter\newcommand{\ps@icrc}{
\renewcommand{\@oddhead}{Proc.~\icrc, \slshape{OG 3.2.18}\hfil}}
\makeatother\thispagestyle{icrc}
%
%

\begin{center}
{\LARGE \bf 
The GALPROP program for cosmic-ray propagation:\\ new developments}
\end{center}

\begin{center}
{\bf Andrew W.~Strong$^{1}$ and Igor V.~Moskalenko$^{1,2}$}\\
{\it $^{1}$MPI f\"ur extraterrestrische Physik, D--85740 Garching, Germany\\
$^{2}$Institute for Nuclear Physics, Moscow State University, 119 899 Moscow, 
   Russia}
\end{center}

\begin{center}
{\large \bf Abstract\\}
\end{center}
\vspace{-0.5ex}
The cosmic-ray propagation code GALPROP has been generalized to include
fragmentation networks of arbitrary complexity.  The code can now
provide an alternative to leaky-box calculations for full isotopic
abundance calculations and has the advantage of including the spatial
dimension which is essential for radioactive nuclei.  Preliminary
predictions  for \Ferat, \Berat, \Alrat, \Clrat, and \Mnrat\ are
presented in anticipation of new experimental isotopic data.
%

\vspace{1ex}

%
%

\section{Introduction:}

The GALPROP program (Strong and Moskalenko 1998, Moskalenko and Strong
1998, Strong et al.\ 1999) performs cosmic-ray propagation calculations
for nuclei, electrons and positrons and computes \gray and
synchrotron emission in the same framework.  The 3D spatial approach with a realistic distribution of interstellar gas
distinguishes it from leaky-box calculations.  Originally we considered
that the complexity of the spatial model would preclude using
cosmic-ray reaction networks with more than a few species, and our
results so far have hence been obtained using a simplifying `weighted
cross sections' method.  However we have now been able to generalize
the scheme to include reaction networks of arbitrary complexity.  The
scheme can hence potentially compete with leaky-box calculations for
computation of isotopic abundances, while retaining  the spatial
component essential for radioactive nuclei, electrons and gamma rays.
Even for stable nuclei it has the advantage of a physically-based
propagation scheme with a spatial distribution of sources rather than
an {\it ad hoc} path length distribution.  It also facilitates tests
of reacceleration.  Here we give a progress report on the new extension
of the method  and present some illustrative results. The technique
should be of interest for the interpretation of new measurements such
as those becoming available from ACE and ISOMAX.

\section{Method:}

GALPROP solves the Galactic CR propagation equation numerically on a grid in 3D
with cylindrical symmetry. In the original simplified approach for
nuclei we propagated first the primaries, then the source function for
each secondary was computed and then the secondaries were propagated.
Multiple progenitors and tertiary etc. reactions were handled by
weighting the cross-sections according to the observed abundances,
which was clearly a compromise solution.

The new extended scheme instead handles the reaction network
explicitly, as follows:

1. Propagate all primary species from an assumed set of source abundances;

2. Compute the resulting spallation source function for all species;

3. Propagate all species  using the sum of primary and spallation sources;

4. Iterate steps 2 and 3 until converged.

After the second iteration the result is already accurate for the pure
secondary component, after the third iteration it is accurate for
tertiaries, and so on.  Hence in practice only 3 or 4 such iterations
are necessary for a complete solution of the network.  The method is
more time-consuming than the simple approach since many more species
are included and because of the several iterations required.

\section{Results:}

We have applied the method to a network of 87 nuclei from protons to
Ni, including explicitly all stable species and radioactive species
with half-life more than $10^5$ years.  The network of channels for
decayed cross-sections is based on Nuclear Data Sheets.  Isotopic cross
sections are based on measured values where available, up to and
including Webber et al.\ (1998). For cases where  measurements at
enough energies exist we interpolate a smooth function.  Otherwise we
use the Webber et al.\ (1990) cross-section code, renormalizing to
measurements where they exist.  The interstellar
He abundance is taken as 0.11 by number.  Source abundances are from
DuVernois and Thayer (1996), with solar isotopic ratios within a given
element. Propagation parameters for this model with reacceleration
are:  halo size $z_h = 4$ kpc, $v_A = 20$ \vunits, diffusion coeffient
$D=D_0\beta{(p/p_0)}^{1\over3}$, where $D_0 = 6.75$\Dunits, $p_0 = 3$
GeV.  The diffusion coefficient, adjusted to fit \BC, differs slighly
from that used in the original work for the same $z_h$ (Strong et
al.\ 1998: $D_0 = 6$\Dunits), reflecting the more detailed treatment
and updated cross sections.

\begin{figure}[thb]
\psfig{file=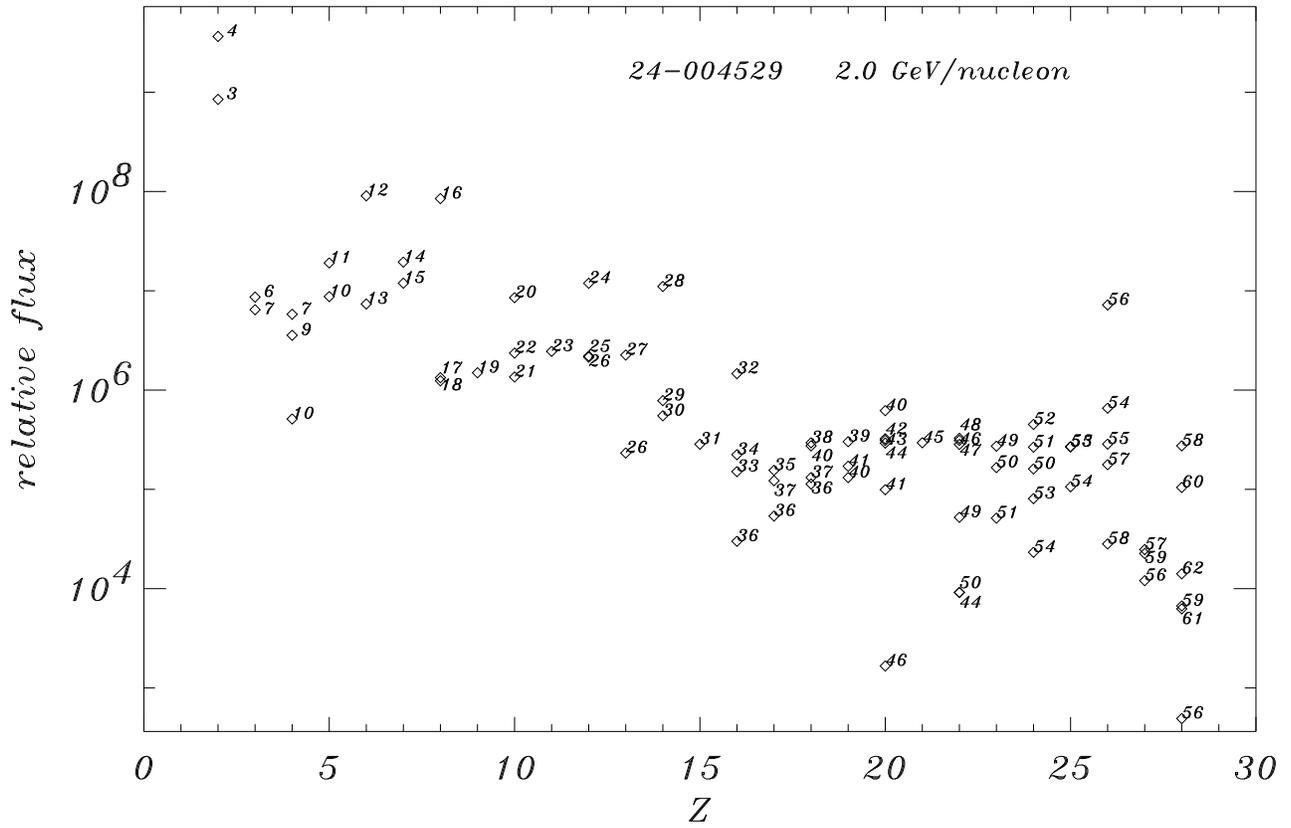,width=\fwa,height=\fha,clip=}
\caption[]{ Model isotopic fluxes at 2 GeV/nucleon as function of $Z$;
individual $A$ values are marked. NB: This plot is illustrative of the
method but not to be taken as predictions for evaluation purposes.
\label{fig1} }
\end{figure}

Fig.~\ref{fig1} shows computed fluxes of all the included isotopes at 2
GeV/nucleon.   This result is illustrative of the method but not to be
taken as predictions for evaluation purposes.

\begin{figure}[thb]
\psfig{file=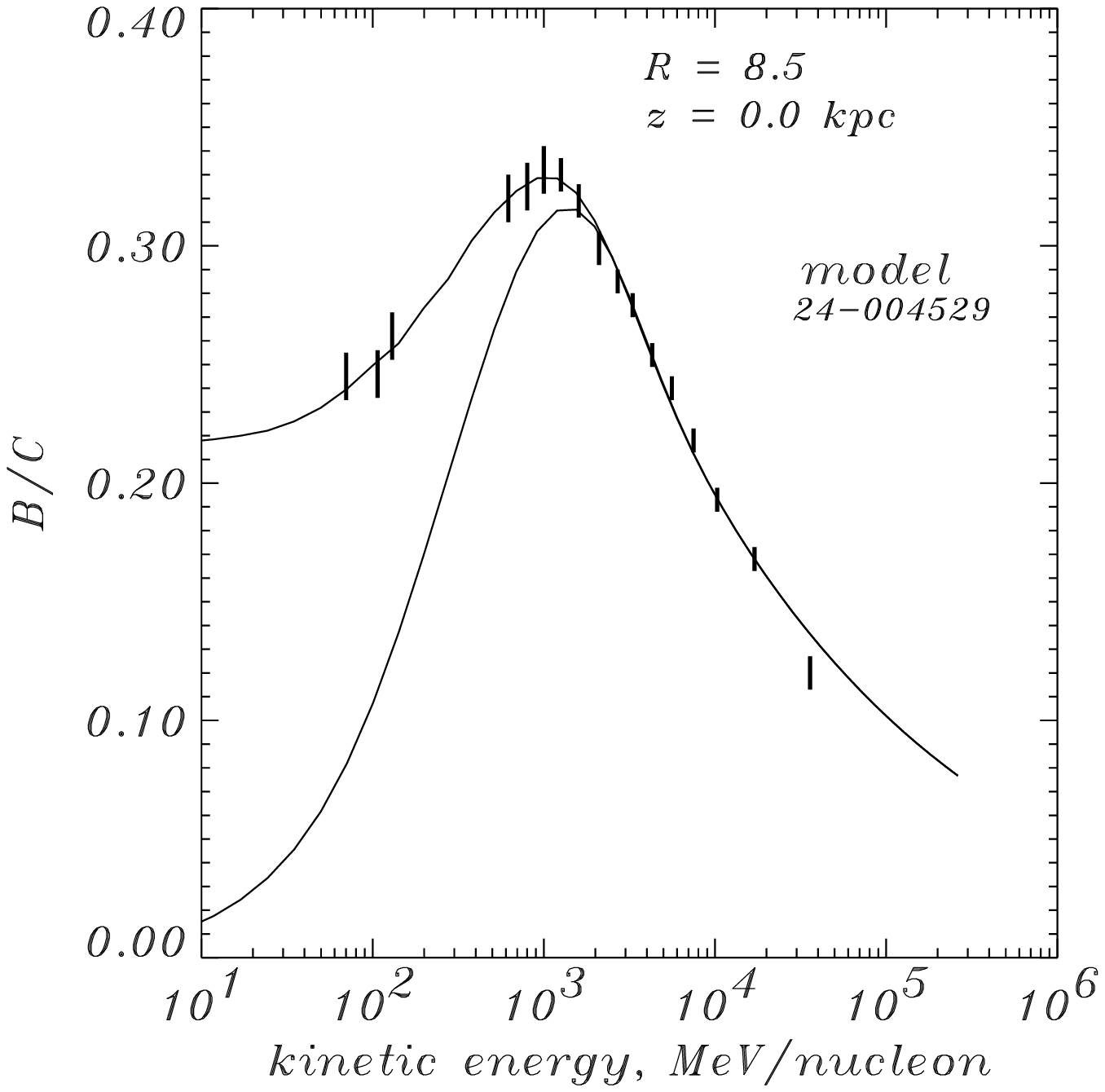,width=\fwb,clip=}
\psfig{file=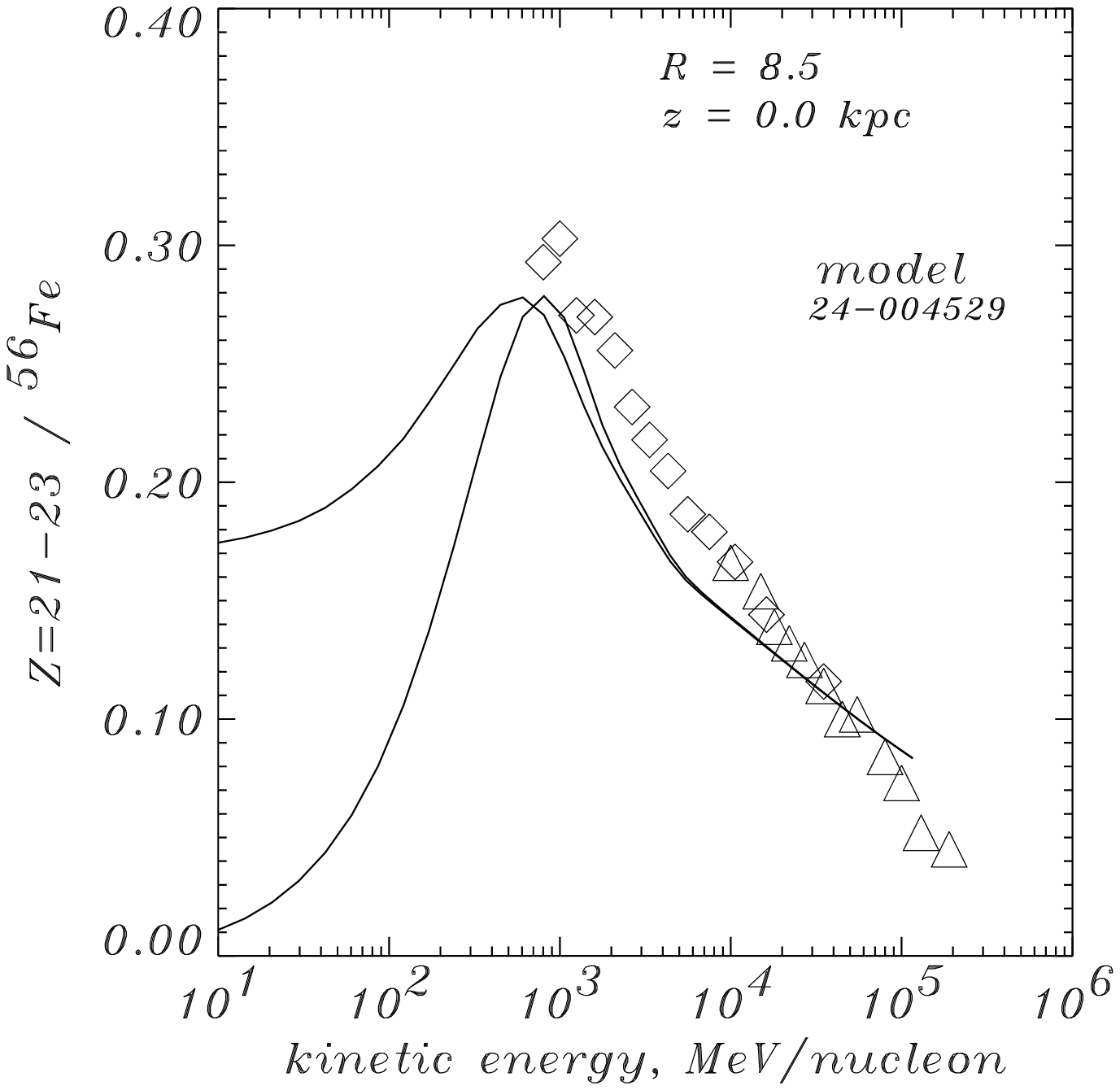,width=\fwb,clip=}
\caption[]{{\it Left}: \BC\ interstellar and modulated to 500 MV for
diffusive reacceleration model with $z_h$ = 4 kpc.  Data compilation:
Webber et al.\ (1996).
{\it Right}: The same for \Ferat\ $(Z=21-23/\,^{56}$Fe).  Data:
\odiamond -- Engelmann et al.\ (1990), \otriangle -- Binns et
al.\ (1988).
\label{fig2} }
\end{figure}

\begin{figure}[thb]
\psfig{file=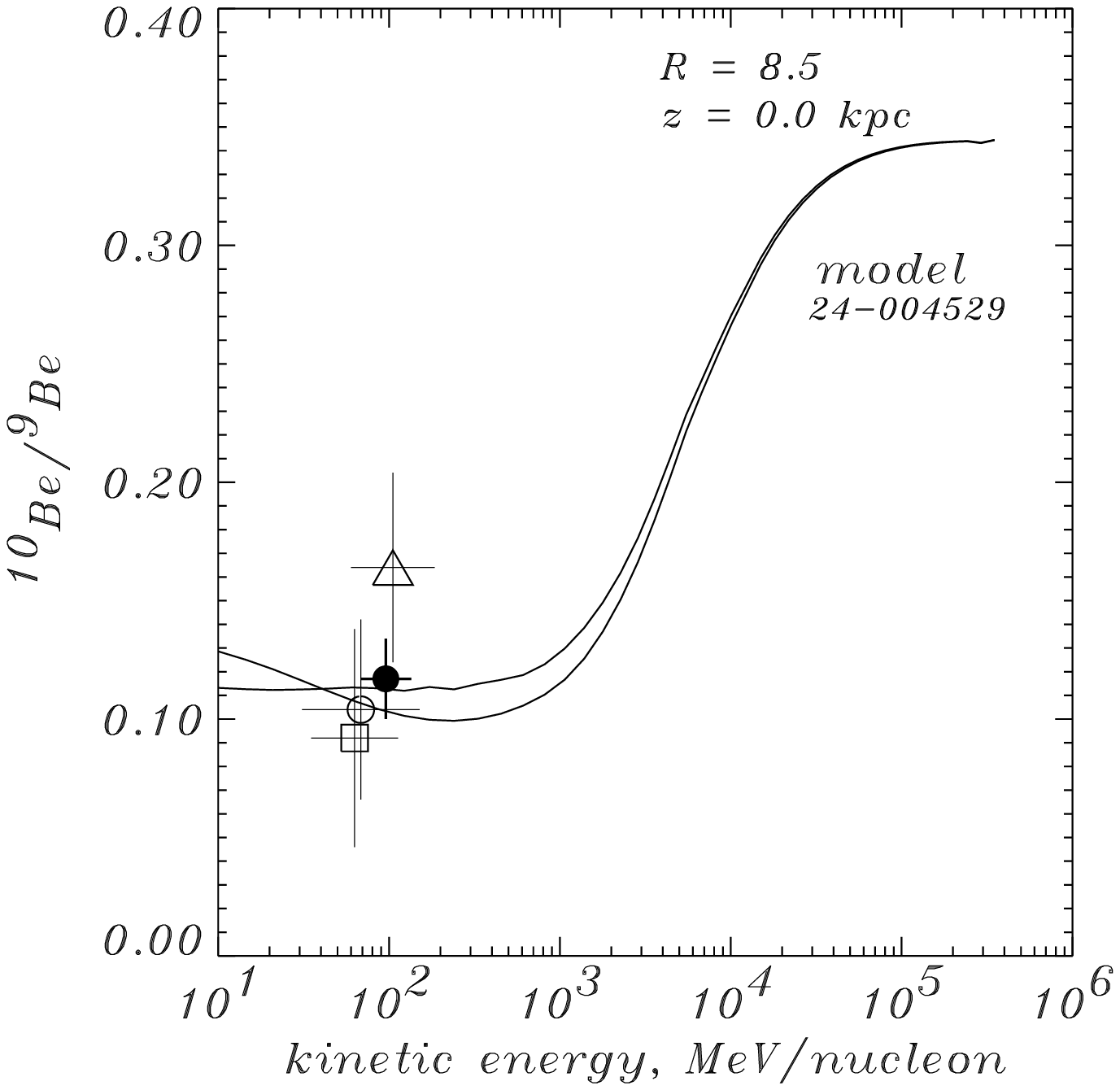,width=\fwb,clip=}
\psfig{file=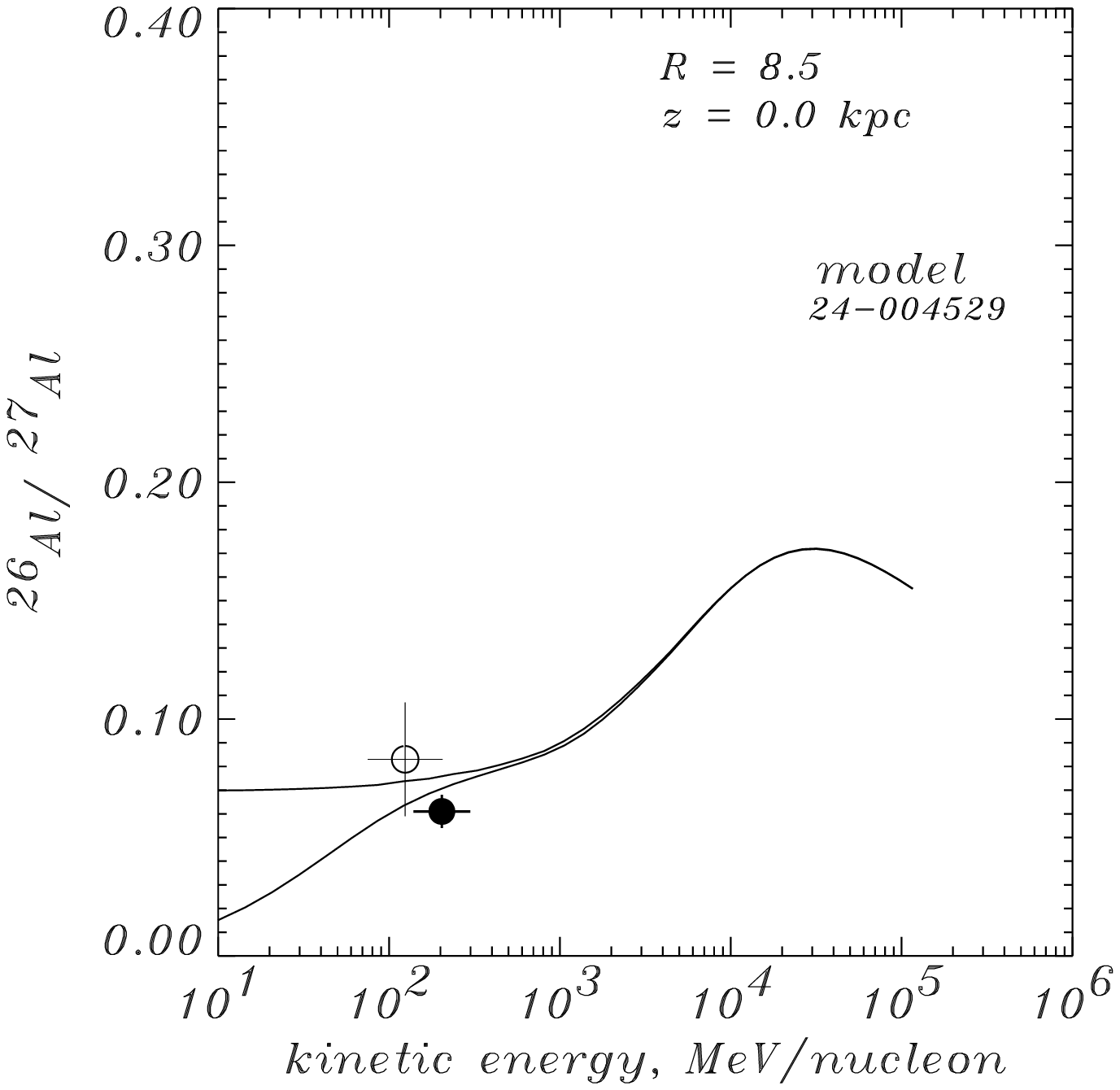,width=\fwb,clip=}
\caption[]{ 
Model interstellar and modulated (500 MV) ratios.  
{\it Left}: \Berat. Data from Lukasiak et al.\ (1994a) (\obox --
Voyager--1,2, \ocircle -- IMP--7/8, \otriangle -- ISEE--3) and Connell
(1998) (\fcircle -- Ulysses).
{\it Right}: \Alrat. Data: \ocircle -- Lukasiak et al.\ (1994b),
\fcircle -- Simpson and Connell (1998).  Note that the data points
shown are at the measured (not interstellar) energies.
\label{fig3} }
\end{figure}

\begin{figure}[thb]
\psfig{file=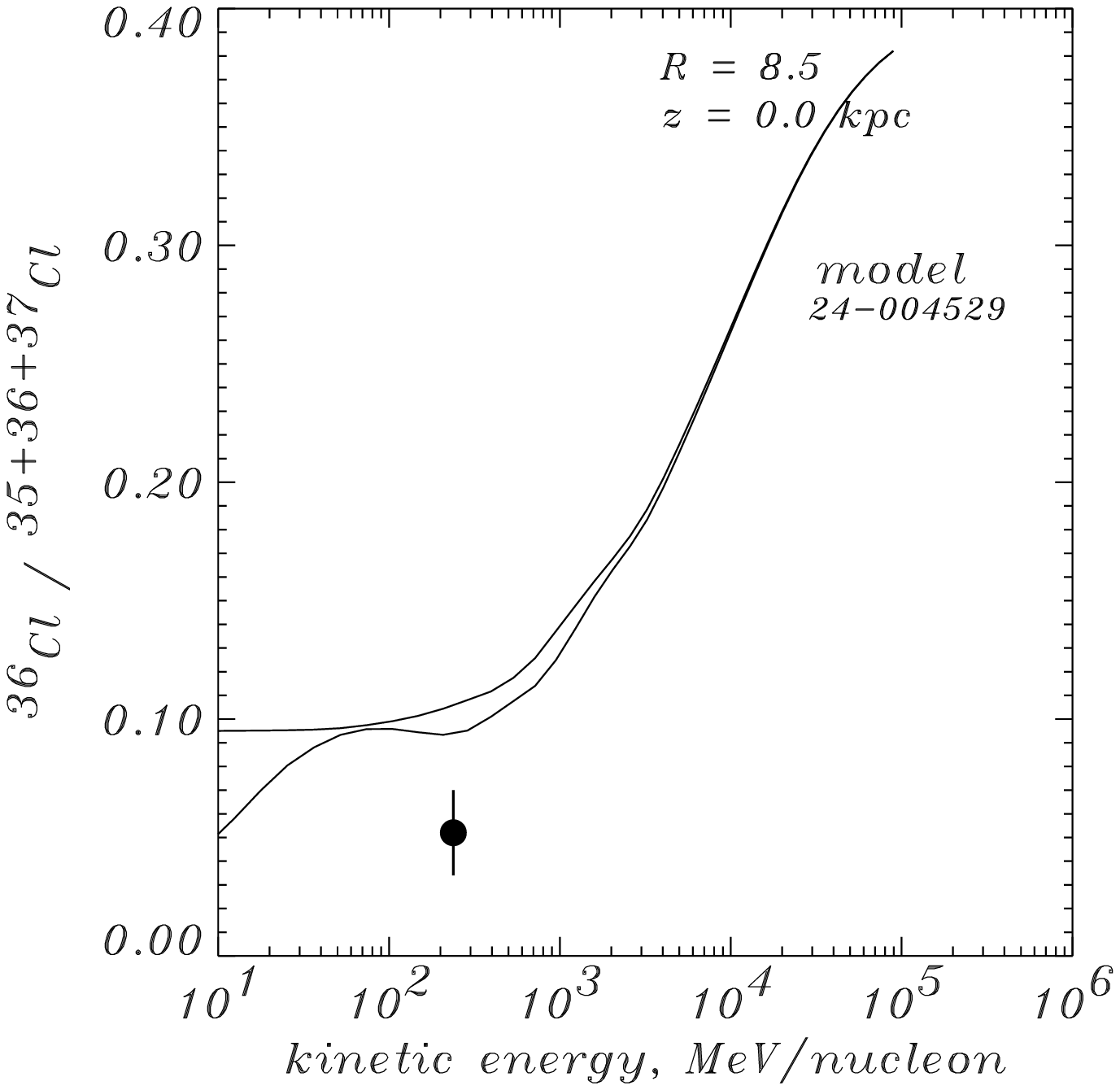,width=\fwb,clip=}
\psfig{file=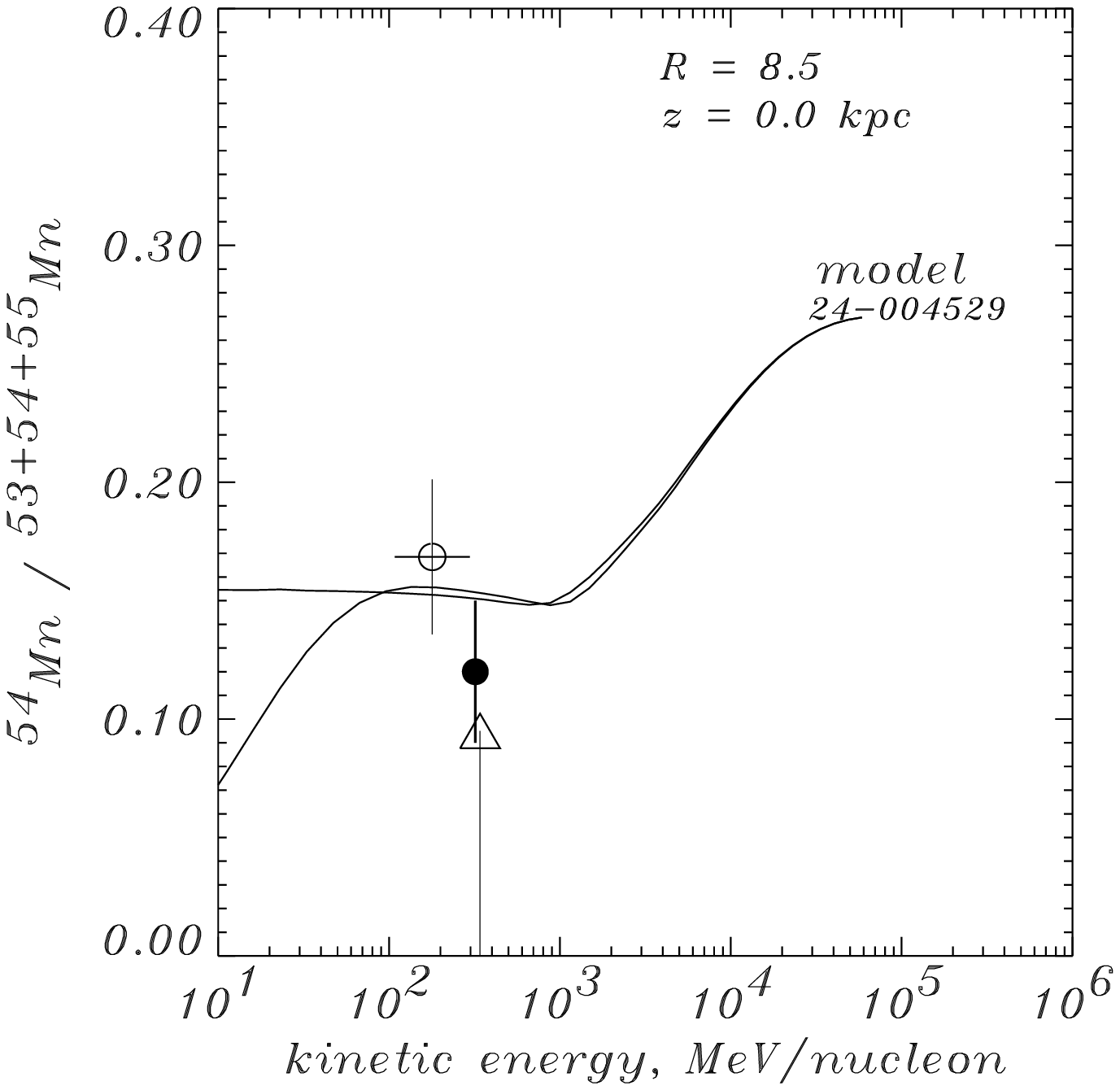,width=\fwb,clip=}
\caption[]{Model interstellar and modulated (500 MV) ratios.
{\it Left}: \Clrat. Data: Connell et al.\ (1998); \\ 
{\it Right}:  \Mnrat. Data: \fcircle -- Duvernois (1997), \ocircle --
Lukasiak et al.\ (1995), \otriangle -- Leske (1993).
\label{fig4} }
\end{figure}

Fig.~\ref{fig2} shows \BC\ and  \Ferat.  This model with reacceleration
reproduces the sub-Fe data reasonably well considering that the model
was adjusted only to fit B/C.  The deviation from the observed \Ferat\
shape is however noticeable, as discussed by Webber (1997).
Figs.~\ref{fig3} and \ref{fig4} show the radioactive isotope ratios
\Berat, \Alrat, \Clrat, and \Mnrat\ (using an estimated $^{54}$Mn
halflife of $6.3 \times 10^5$ years, Wuosmaa et al.\ 1998).  Modulation
is for nominal values of the modulation parameter but this has little
effect on the comparison due to the small energy dependence of these 4
ratios in the 100--1000 MeV/nucleon range.  In Strong and Moskalenko
(1998) we obtained a range 4 kpc$<z_h<$12 kpc for the halo height; in
the present improved model the \Berat, \Alrat\ predictions are
consistent with this.  A complete grid of models will be generated in
future to obtain  new limits.  Our \Clrat\ and \Mnrat\ predictions are
preliminary and do not yet provide significant constraints.

\section{Conclusions:}

The extension of GALPROP to complete networks of nuclei has been
accomplished, but further work is needed especially on the
cross-sections before the method can be used for detailed evaluation.
As usual the software will be made
available on the WWW at http://www.gamma.mpe--garching.mpg.de/$\sim$aws/aws.html
\vspace{1ex}
\begin{center}
{\Large\bf References}
\end{center}
Binns, W.R., et al. 1988, \apj 324, 1106\\ 
Connell, J.J.  1998, \apj 501, L59\\ 
Connell, J.J. et al. 1998, \apj 509, L97\\ 
DuVernois, M.A. 1997, \apj 481, 241 \\ 
DuVernois, M.A., \& Thayer, M.A. 1996, \apj 465, 982\\
Engelmann, J.J., et al. 1990, \aap 233, 96\\
Leske, R.A. 1993, \apj 405, 567\\ 
Lukasiak, A., et al.  1994a, \apj 423, 426\\
Lukasiak, A., et al.  1994b, \apj 430, L69\\
Lukasiak, A., et al.  1995, Proc.\ 24th ICRC (Roma) 2, 576 \\ 
Moskalenko, I.V., \& Strong, A.W.  1998, \apj 493, 694\\
Simpson, J.A., \& Connell, J.J.  1998, \apj 497, L85\\
Strong, A.W., \& Moskalenko, I.V.  1998, \apj 509, 212\\
Strong, A.W., \& Moskalenko, I.V., \& Reimer, O.  1999, submitted
  (astro--ph/9811296)\par (see also these proceedings OG 2.4.03)\\
Webber, W.R., Kish, J.C., \& Schrier, D.A. 1990, Phys.\ Rev.\ C 41, 566\\
Webber, W.R., et al.  1996, \apj 457, 435\\
Webber, W.R.  1997, Spa.\ Sci.\ Rev.\ 81, 107 \\
Webber, W.R., et al.  1998, \apj 508, 940; {\it ibid.}, p.949\\
Wuosmaa, A.H., et al. 1998, Phys.\ Rev.\ Lett.\ 80, 2085

\end{document}